\documentclass[preprint]{aastex}

\shorttitle{IR Spectra of FUors}
\shortauthors{Greene, Aspin, \& Reipurth}

\begin{document}

\title{High Resolution Near-Infrared Spectroscopy of FUors and
FUor-like stars\footnote{Much of the data presented herein were
obtained at the W.M. Keck Observatory from telescope time allocated to
the National Aeronautics and Space Administration through the agency's
scientific partnership with the California Institute of Technology and
the University of California.  The Observatory was made possible by
the generous financial support of the W.M. Keck Foundation.}}


\author{Thomas P. Greene\altaffilmark{2,3}}
\email{thomas.p.greene@nasa.gov}

\author{Colin Aspin\altaffilmark{3,4}}
\email{caa@ifa.hawaii.edu}

\author{Bo Reipurth\altaffilmark{4}}
\email{reipurth@ifa.hawaii.edu}

\altaffiltext{2}{NASA Ames Research Center, M.S. 245-6, Moffett
Field, CA 94035-1000}

\altaffiltext{3}{Visiting Astronomer at the Infrared Telescope
Facility which is operated by the University of Hawaii under contract
to the National Aeronautics and Space Administration.}

\altaffiltext{4}{Institute for Astronomy, University of Hawaii, 640 N.
A`ohoku Place, Hilo, HI 96720}

\begin{abstract}
We present new high resolution ($R \simeq 18,000$) near-infrared
spectroscopic observations of a sample of classical FU~Orionis stars
(FUors) and other young stars with FUor characteristics that are
sources of Herbig-Haro flows.  Spectra are presented for the region
$\lambda =$ 2.203 -- 2.236~$\mu$m which is rich in absorption lines
sensitive to both effective temperatures and surface gravities of
stars.  Both FUors and FUor-like stars show numerous broad and weak
unidentified spectral features in this region.  Spectra of the 2.280
-- 2.300~$\mu$m region are also presented, with the 2.2935~$\mu$m
$v$=2--0~CO absorption bandhead being clearly the strongest feature
seen in the spectra all FUors and Fuor-like stars.  A
cross-correlation analysis shows that FUor and FUor-like spectra in
the 2.203 -- 2.236~$\mu$m region are not consistent with late-type
dwarfs, giants, nor embedded protostars.  The cross-correlations also
show that the observed FUor-like Herbig-Haro energy sources have
spectra that are substantively similar to those of FUors.  Both object
groups also have similar near-infrared colors.  The large line widths
and double-peaked nature of the spectra of the FUor-like stars are
consistent with the established accretion disk model for FUors, also
consistent with their near-infrared colors.  It appears that young
stars with FUor-like characteristics may be more common than projected
from the relatively few known classical FUors.

\end{abstract}


\section{Introduction}

FU~Orionis stars (FUors) are rare.  Classical FUors have been
identified by large ($\sim 5$ mag) increases in brightness and
luminosity followed by fading over decades, and only a few young stars
have been confirmed as such, notably FU Ori, V1057 Cyg, V1515
Cyg, V1735 Cyg \citep{H77}, V346 Nor \citep{GF85}, and V733 Cep
\citep{RABBCH07}.  Additionally, while in outburst, they exhibit
optical and near-IR spectra similar to FU~Orionis itself \citep{H77,
HK96, HPD03}.  The absorption line widths (interpreted as rotational
velocities) and derived spectral types of FUors change with
wavelength, varying from F or G in the visible to M type in the
near-IR. Their spectra also indicate supergiant or giant surface
gravities.  These characteristics have been modeled as arising from
young stars with massive accretion disks, the latter dominating the
spectra and luminosities of these objects \citep[see][]{HK96}.  The
accretion disk model has been quite successful in explaining the basic
visible-to-IR spectral features and energy distributions of FUors
\citep{HK85, KHH88, HHC04, GHCWIFSF06}, although in detail there are
still discrepancies \citep{HPD03}.  The only published (and modeled)
high resolution near-IR spectra of FUors have been in the vicinity of
the first vibrational overtone $v = 0 - 2$ of CO ($\lambda \gtrsim
2.294 \mu$m), providing temperature and rotation information.  New
observations at other near-IR wavelengths may aid in determining the
range of effective temperatures and surface gravities over which FUor
spectral features arise in the near-IR, providing more constraints on
applicable physical models.

The general symmetry and relatively even spacing often seen in
Herbig-Haro object knots strongly suggest that they are somehow
linked to episodic outbursts from their parent star plus disk systems
\citep{D78,R89}.  This suggests that the sources of HH objects and
FUor outbursts may possibly be related.  \citet{RA97} found that five
young stars associated with HH flows (termed Herbig-Haro energy
sources or HHENs) had low resolution K-band spectra that are very
similar to those of FUors.  If episodic outbursts do generally
indicate FUor activity, then there may be many more young stars with
FUor characteristics than the number predicted by extrapolating
numbers from the few known classical FUors.  In the following
discussions, we use the term FUor-like star to indicate objects that
have spectral similarities to the classical FUors, but for which no
eruption has been witnessed \citep{RHKSB02}.

How similar are classical FUors and FUor-like stars?  Both are young
stars that are likely undergoing episodes of high accretion.  However,
their luminosities may be powered by different physical processes or
they may be at different phases of their outburst and decay cycles.
It is also important to understand how similar these objects are to
embedded (Class~I) protostars accreting at fairly high rates (e.g.,
$\dot{M} \sim 10^{-6} M_{\odot}$ yr$^{-1}$) and to investigate whether
there are any spectral similarities between protostars, FUors, and
FUor-like objects.  Also, do any of these objects show spectral
characteristics similar to other young eruptive variables such as
EX~Lupi-type (EXors) young stars?  EXors undergo multiple short
duration optical outbursts of 1 -- 5~mag but have K or M dwarf optical
spectra dominated by emission lines, much different from FUor spectra
\citep[e.g., see][]{HAGI01}.

We believe that such fundamental questions can be addressed with new
high resolution spectra.  Such spectra must cover near-IR wavelengths
since many FUor-like stars are highly extinguished and cannot be
observed in visible light.  The low resolution near-IR spectra of
\citet{RA97} did show strong similarities between FUors and FUor-like
objects, but those spectra were also consistent with giant stars.  One
object with such a near-IR spectrum in close proximity to a young star
was recently found to be a background giant star \citep{AG07}.
Linking other young stars to FUors will improve our knowledge of the
number and types of young stars with high accretion rates, and
detailed modeling may reveal more information on physical structures
and accretion mechanisms or processes.

We have conducted a new near-IR spectroscopic study of FUors and
FUor-like stars covering a relatively broad range of wavelengths near
2~$\mu$m, including spectral features sensitive to both effective
temperature and surface gravity.  We present these new data in $\S 2.$
The properties of the individual FUor-like stars are discussed in $\S
3$.  In $\S 4$ we analyze the similarity of the spectra of FUor-like
stars to those of {\it i)} classical FUors, {\it ii)} late-type
spectral standards, and {\it iii)} Class~I protostars.  We discuss the
likely physical similarities of these stars and the possible origins
of their attributes in $\S 5$.

\section{Observations and Data Reduction}

High resolution near-IR spectra of several classical FUors and
FUor-like stars were acquired with the NASA IRTF, Keck~II, and
GEMINI-South telescopes.  The observation dates, total integration
times, signal-to-noise, equipment used, and coordinates of all
observed objects are given in Table~\ref{tbl-1}.

\subsection{Object Sample}

The observed FUors and FUor-like objects are listed in
Table~\ref{tbl-1}.  The FUor-like Herbig-Haro sources V883~Ori,
HH~354~IRS, HH~381~IRS, and L1551~IRS~5 were observed by \citet{RA97}
who found that their low resolution ($R \sim 420$) $K$-band spectra
were similar to those of FUors.  Parsamian~21 has not to our
knowledge been observed previously with near-infrared spectroscopy.
The spectra of L1551~IRS~5 were previously published in
\citet{DGCL05}.  New and unique observations were made of the FUor
V1057~Cyg in the 2.2935~$\mu$m CO bandhead region, and we use its
previously published spectrum in the 2.2075~$\mu$m Na line region
\citep{GL97}. New spectra of FU~Ori itself were also acquired.

\subsection{IRTF Observations}

Near-IR spectra of V1057~Cyg were acquired on UT 1999 August 30
January with the 3.0~m NASA Infrared Telescope Facility on Mauna Kea,
Hawaii, using the CSHELL facility single-order cryogenic echelle
spectrograph \citep{TTCHE90,GTTC93}.  Spectra were acquired with a
1\farcs0 (5-pixel) wide slit on the dates indicated in
Table~\ref{tbl-1}.  This provided a spectroscopic resolution $R \equiv
\lambda / \delta \lambda$ = 21,000 (14~km~s$^{-1}$).  The spectrograph
was fitted with a 256~$\times$~256 pixel InSb detector array, and
custom circular variable filters (CVFs) manufactured by Optical
Coating Laboratories Incorporated were used for order sorting.  These
filters successfully eliminated the significant interference fringing
normally produced in CSHELL and other echelle spectrographs which use
CVFs for order sorting.  The plate scale was 0\farcs20 pixel$^{-1}$
along the 30$\arcsec$ long slit (oriented east -- west on the sky),
and all spectra were acquired at a central wavelength setting of
2.29353~$\mu$m corresponding to the $v$=2--0 CO bandhead.  Each
exposure had a spectral range $\Delta \lambda \simeq \lambda / 400$
($\Delta v \simeq$ 700 km s$^{-1}$).

Data were acquired in pairs of exposures of up to 180~s duration
each, with the telescope nodded $10\arcsec$ east or west between
exposures so that object spectra were acquired in all exposures.  The
total integration time on V1057~Cyg was 13.0 minutes.  The A1V star
HR~8585 was observed at nearly identical airmass for telluric
corrections.  Spectra of the internal CSHELL continuum lamp were taken
for flat fields, and exposures of the internal CSHELL Ar and Kr lamps
were used for wavelength calibrations.

\subsection{Keck Observations}

Spectra of FU~Ori and all five FUor-like stars were acquired on
UT 2001 July 7--8, 2001 November 6, and 2007 March 6. These
data were obtained with the 10-m Keck~II telescope on Mauna Kea,
Hawaii, using the NIRSPEC multi-order cryogenic echelle facility
spectrograph \citep{McLeanetal98}.  Spectra were acquired with a
0\farcs58 (4-pixel) wide slit, providing spectroscopic resolution $R
\equiv \lambda / \delta \lambda$ = 18,000 (16.7~km~s$^{-1}$).  The
plate scale was 0\farcs20 pixel$^{-1}$ along the 12$\arcsec$ slit
length, and the seeing was typically 0\farcs5--0\farcs6.  The NIRSPEC
gratings were oriented to allow orders containing the 2.1066~$\mu$m
Mg and 2.1099~$\mu$m Al lines, the 2.1661~$\mu$m HI Br~$\gamma$ line,
the 2.206 and 2.209~$\mu$m Na lines, and the 2.2935~$\mu$m CO
bandhead regions to fall onto the instrument's 1024 $\times$ 1024
pixel InSb detector array.  The NIRSPEC-7 blocking filter was used to
image these orders on the detector.  NIRSPEC was configured to
acquire simultaneously multiple cross-dispersed echelle orders 32--36
(2.08--2.37~$\mu$m, non-continuous) for all objects.  Some objects
were also observed in adjacent orders, resulting in a somewhat
increased spectral range at longer or shorter wavelengths.  Each
order had an observed spectral range $\Delta \lambda \simeq \lambda /
67$ ($\Delta v \simeq$ 4450 km~s$^{-1}$).

The slit was held physically stationary during the exposures
and thus rotated on the sky as the non-equatorially-mounted
telescope tracked when observing.  Data were acquired in pairs of
exposures of durations from 5--500~s each, with the telescope nodded
$6\arcsec$ along the slit between frames so that object spectra were
acquired in all exposures. Early-type (B9--A2) dwarfs were observed
for telluric correction of the FUor and FUor-like spectra.  The
telescope was automatically guided with frequent images from the
NIRSPEC internal ``SCAM'' IR camera during all exposures of more than
several seconds duration.  Spectra of the internal NIRSPEC continuum
lamp were taken for flat fields, and exposures of the Ar, Ne, Kr, and
Xe lamps were used for wavelength calibrations.

\subsection{GEMINI Observations}

Spectra of FU~Ori were acquired on UT 2006 April 03 with the
Phoenix near-IR spectrograph \citep{Hinkle02} on the 8-m GEMINI-South
telescope on Cerro Pachon, Chile.  Spectra were acquired with a
$0\farcs35$ (4-pixel) wide slit, providing spectroscopic resolution
$R \equiv \lambda / \delta \lambda$ = 40,000 (7.5~km~s$^{-1}$).  The
grating was oriented to observe the spectral range $\lambda$ =
2.2194--2.2290~$\mu$m in a single long-slit spectral order, and a
slit position angle of $90^{\circ}$ was used.  Data were acquired in
a pair of exposures of 120 s duration each, with the telescope nodded
$5\arcsec$ along the slit between frames so that object spectra were
acquired in all exposures.  The B0.5 star HR~4730 was observed for
telluric correction of the spectra.  Observations of a Th-Ar-Ne
hollow-cathode lamp were used for wavelength calibrations, and
continuum lamp observations were used for flat fields.

\subsection{Data Reduction}

All data were reduced with IRAF. First, object and sky frames were
differenced and then divided by normalized flat fields.  Next,
bad pixels were fixed via interpolation, and spectra were extracted
with the APALL task.  Spectra were wavelength calibrated using
low-order fits to lines in the arc lamp exposures, and spectra at
each slit position of each object were co-added.  Instrumental and
atmospheric features were removed by dividing wavelength-calibrated
object spectra by spectra of early-type stars observed at similar
airmass at each slit position. Final spectra were produced by
combining the spectra of both slit positions for each object and then
normalizing them so that they had a mean relative flux of 1.0 in each
order.

\section{Notes on Individual FUor-like Objects}

We now discuss what is already known about the individual FUor-like
HHENs, with a focus on the properties that indicate episodic
variability or other properties that indicate FUor-like behavior.

\subsection{L1551~IRS~5}

L1551~IRS~5 was discovered in the near-IR survey of the Taurus cloud
by \citet{SSV76}.  It was found to be associated with a molecular
outflow \citep{SLP80} and is a close binary system
\citep[e.g.,][]{BC85}.  \citet{RCCLRT03} discovered that both binary
components drive aligned ionized bipolar jets from cm wavelength
observations.  L1551~IRS~5 was confirmed as a FUor-like star by
\citet{SS93} after the initial classification by \citet{MSSSA85}.  It
is an IRAS source (04287+1801) with a 12~$\mu$m flux of $\sim$10~Jy and
was later found to be a triple system with separations 47 and 13~AU
\citep{LT06}.  It is the driving source of an optical bipolar jet
designated HH~154 \citep{MF83}.

\subsection{V883~Ori}

This object was first noted as a faint star illuminating an extensive
reflection nebula (designated IC~430) on photographic plates dating
from 1888.  In the H$\alpha$ emission line survey of \citet{Haro53},
it was described as being faint with some nebulosity and given the
designation Haro~13a.  Optical spectroscopy of the reflection
nebulosity suggested that V883~Ori was a FUor \citep{SS93}.  V883~Ori
possesses a curving tail of nebulosity, as do many FUors, and is an
IRAS source (05358-0704) with a 12~$\mu$m flux of 52~Jy.  It has a
bolometric luminosity of $\sim$400~L$_{\odot}$ and sub-mm observations
have determined it has an (unresolved) circumstellar gas+dust mass of
$\sim$0.4~M$_{\odot}$ \citep{DMW88}.  It is thought to be the driving
source of HH~183 \citep{SSWMW86}.

\subsection{Parsamian~21}

Parsamian~21 was first noted in the catalog of cometary nebulae by
\citet{Par65}.  From optical images and spectroscopy, \citet{SN92}
concluded that this source was a possible FUor and discovered a small
bipolar HH flow (later labeled as HH~221) emanating from the star.
\citet{Ketal07} found that the H$\alpha$ knots first observed by
\citet{SN92} were moving at velocities 120 -- 500 km s$^{-1}$, and
their high resolution near-IR direct and polarimetric images reveal a
circumstellar envelope, a polar cavity, and an edge-on disk.
Parsamian~21 is associated with a cold IRAS source (19266+0932) with
12~$\mu$m flux 0.8~Jy and a 100~$\mu$m flux of 15~Jy.  \citet{BL83}
found no associated CO outflow from the star.

\subsection{HH~381~IRS}

HH~381~IRS shows prominent optical \citep{DRB97} and near-IR
nebulosity \citep{CRT07} and is an IRAS source (20658+5217) with a
12~$\mu$m flux of 0.3~Jy and 100~$\mu$m flux of 11~Jy.  \citet{DRB97}
also considered it the driving source of the HH objects HH~381/382.
HH381-IRS was found to have a near-IR spectrum almost identical to
that of L1551~IRS~5 by \citet{RA97}.  \citet{JENAM-2007} found that
the star and associated nebulosity were very faint on DSS-1 (1953)
plates while bright in DSS-2 (1990) images, and even brighter on
recent CCD images.  Finally, it is not yet conclusive that HH~381 (and
in fact HH~380, and 382) originates from HH381-IRS since there are
several other IRAS sources in the region.

\subsection{HH~354~IRS}

HH354-IRS is located in Lynds~1165 and drives a large-scale HH flow
covering ~$\sim$2.4~pc \citep{RBD97}.  It is associated with an IRAS
source (22051+5848) and has a weak 12~$\mu$m flux of 0.3~Jy and a
strong 100~$\mu$m flux of 94~Jy.  Its bolometric luminosity is
$\sim$120~L$_{\odot}$.  It was found to have a near-IR spectrum very
similar to L1551~IRS~5 \citep{RA97} and possess a strong sub-mm flux
suggesting a gas+dust mass of $\sim$~30~M$_{\odot}$ and a CO outflow
\citep{VRC02}.

\section{Analysis and Results}


Spectra of the classical FUors V1057~Cyg and FU~Ori are shown over the
2.203- 2.236~$\mu$m and the 2.280--2.300~$\mu$m ranges in
Figure~\ref{fig1}.  The GEMINI-South spectrum of FU~Ori is not shown
here because it was superseded by the shown Keck ~II NIRSPEC spectrum.
However, the GEMINI data are used in the subsequent analysis.
Figure~\ref{fig1} also shows the spectra of the M4V star GJ~402
\citep[from][]{DGCL05}, the M2~Iab star $\alpha$~Ori
\citep[from][]{WH96}, and the veiled ($r_k$ = 1.7) K5--7 Class~I
protostar $\rho$~Oph~IRS~63 \citep[from][]{DGCL05}.  The spectra of
$\alpha$~Ori were smoothed so that the CO bandhead has approximately
the same slope as the FUors, requiring an effective broadening of
$v$~sin~$i \simeq 100$ km s$^{-1}$.  The $\alpha$~Ori spectra were
also artificially veiled by $r_k$ = 1.0 so that their features have
depths similar to those found in the FUors and FUor-like stars.  The
spectra of the five FUor-like HHENs and $\alpha$~Ori are shown in
Figure~\ref{fig2}.

\subsection{Spectral Properties}

The M4~V star GJ~402 shows strong neutral atomic absorption lines of
Na, Sc, Ti, Fe, and Mg species that are diagnostic of both
effective temperature and surface gravity \citep[e.g.,
see][]{DGCL05}.  The $v$=2--0~CO bandhead and red-ward
vibration-rotation lines are also very prominent, and their depths
are good indicators of surface gravity when interpreted together with
the atomic features.  The narrow profiles of all lines (except the
somewhat gravity-broadened Na lines) are indicative of slow,
unresolved rotation ($v$~sin~$i < 17$~km~s$^{-1}$). The Class~I
protostar IRS~63 has similar features but they are much weaker due to
significant near-IR veiling.  They are also broad due to
the object's considerable rotational velocity, $v$~sin~$i =
45$~km~s$^{-1}$ \citep{DGCL05}.  The spectra of the M2~Iab 
star $\alpha$~Ori show significant Na, Ti, and CO absorption
in addition to many relatively weak features \citep[see][]{WH96}.
However, they appear weak and broad in Figure~\ref{fig1} due
to the artificial veiling and rotational broadening applied.

The spectra of the two FUors are clearly different from those of the
stellar standards and the Class~I protostar (Figure~\ref{fig1}).
FU~Ori shows numerous weak and broad absorption lines throughout the
2.280--2.300~$\mu$m range whereas the late-type dwarf, supergiant, and
protostar show discrete features (V1057 Cyg also does not show normal
stellar features over its limited spectral coverage).  The same is true
for the spectra of the five FUor-like stars in Figure~\ref{fig2}.  The
$v$=2--0~CO bandhead and red-ward envelope defining that absorption
band are certainly the most distinguished spectral features in both
the FUors and FUor-like stars.

Optical spectra of FUors have been modeled as being produced in
either a circumstellar disk \citep[e.g.,][]{HK85, KHH88} or else in a
rapidly rotating G-type supergiant atmosphere surrounded by a
chromosphere and a cooler absorbing shell \citep[e.g.,][]{PH92,
HPD03}.  The near-IR spectrum of FU~Ori in the vicinity of the
$\Delta v=2$ CO bandheads and nearby ro-vibration lines has been
successfully modeled as arising in a rotating accretion disk at radii
where the disk photosphere has an effective temperature and surface
gravity characteristic of an M-type giant or supergiant \citep{KHH88,
HHC04}.  Figure~\ref{fig1} shows that the larger bandwidth near-IR
spectra of FU~Ori and V1057~Cyg do share some similarities with the
artificially broadened and veiled spectra of the M2 supergiant
$\alpha$ Ori, but overall they appear to be more similar to the
spectra of the FUor-like stars in Figure~\ref{fig2}.  Likewise, the
FUor-like stars appear to have spectra more similar to FUors than to
rotating, veiled late-type stars or Class I protostars.

\subsection{Cross-Correlation Analysis}

We now quantify the similarity of the spectra of the
FUors and FUor-like stars with those of the stellar dwarfs,
stellar giants, and Class~I protostars by examining their
cross-correlations.  Figure~\ref{fig3} shows the cross-correlation
functions of the 2.203--2.236~$\mu$m spectrum of the FUor-like star
HH~381~IRS against template spectra of FU~Ori (Keck spectrum),
$\alpha$~Ori (M2~Iab), GJ~402 (M4~V), and IRS~63 (late-type
protostar).  Spectra have been shifted slightly to produce
symmetrical cross-correlation peaks at 0~km s$^{-1}$, effectively
eliminating relative radial velocities.  The high central peak and
symmetrical nature of the cross-correlation in the top panel
indicates that the spectra of HH~381~IRS and FU~Ori have very similar
features, but correlations with the giant, dwarf, and embedded
protostar are poor.

The broad cross-correlation peaks in all panels of Figure~\ref{fig3}
indicate that HH~381~IRS has broad spectral features.  The peak of
the cross-correlation with FU~Ori has a half width of approximately
100~km~s$^{-1}$, indicating that the 2~$\mu$m spectra of these
objects are originating from regions rotating with velocities
$v$~sin~$i \sim 70$~km~s$^{-1}$ each (also consistent with the
similarly broad line widths seen in Figures~\ref{fig1} and
\ref{fig2}).  The cross-correlation of HH~381~IRS and the slowly
rotating M4 dwarf GJ~402 is double-peaked, consistent with the
spectrum of HH~381~IRS originating in a rotating disk \citep{KHH88,
HHC04}.  This was also seen in the cross-correlation of HH~381~IRS
(and other FUor-like stars) with the slowly rotating M giant HR~5150
\citep[from][not shown]{DGCL05}.  The cross-correlations with the
other stars do not have double-peaked structures, likely because of
their large rotation velocities (natural in IRS~63 and artificially
added in $\alpha$~Ori).

We have computed cross-correlation functions of all FUor-like stars
over the 2.203--2.236~$\mu$m wavelength interval using FU~Ori (Keck
spectrum), $\alpha$~Ori, HR~5150 (M1.5~III), GJ~402, and the
protostar IRS~63 as templates.  We have computed the \citet{TD79}
$r$-values, a measure of the correlation peak height to the average
noise, for each correlation function.  These values are presented in
Table~\ref{tbl-2}, with values greater than 3 indicating
a significant correlation.

Table~\ref{tbl-2} shows that the spectra of the FUor-like HHENs are
more strongly correlated with FUor spectra than other stars in every
case. Also, all FUor-like stars are significantly correlated with the
spectrum of FU~Ori itself ($r>3$).  The FUor-like stars
L1551~IRS~5, Parsamian~21, HH~354~IRS, and V883~Ori also show nearly
or marginally significant correlations ($r\sim~3$) with one or more
other templates, as does FU~Ori.  This indicates that FUors and
FUor-like stars have some spectral features in common with giants,
dwarfs, or protostars.  However, the much more significant
cross-correlations between FU~Ori and FUor-like stars indicate that
the spectra of these objects are much more similar to each other than
to normal stars or embedded protostars.

In Table~\ref{tbl-3} we compare the cross-correlation results for
each FUor or FUor-like star (over 2.203--2.236~$\mu$m wavelength)
with each other in an attempt to measure the similarities of their
spectral features.  Each object is listed in both columns and rows
and the cross-correlation values are given for each source against
each other source.  Table entries which cross-correlate a source
against itself list infinity ($\infty$) as the value.  This is not
the case, however, for FU~Ori, where we have cross-correlated the
Keck NIRSPEC spectrum taken on UT 2007 Mar 06 (and used in all
other FU~Ori cross-correlations) with the Gemini-S Phoenix spectrum
taken on UT 2006 April 03.

Inspection of the cross-correlation $r-$values shows that the
structure in the spectrum of FU~Ori correlates extremely well with
L1551~IRS~5 ($r\sim$5), V883~Ori ($r\sim$29), Parsamian~21
($r\sim$22), HH~381~IRS ($r\sim$13), and HH~354~IRS ($r\sim$8).  This
implies that all five sources are spectrally very similar to FU~Ori.
The inter-correlations of these five sources suggests that L1551~IRS~5
is most similar to HH~354~IRS ($r\sim$8), V883~Ori is most similar to
FU~Ori ($r\sim$29), Parsamian~21 is most similar to V883~Ori
($r\sim$24), HH~381~IRS is most similar to Parsamian~21 ($r\sim$22),
and HH~354~IRS is most similar to Parsamian~21 ($r\sim$10).  We note
also that all cross-correlation values are above the $r$=3
significance threshold indicating that all have significant
similarities to all others.  The significance of the cross-correlation
of the two different FU~Ori observations is very high ($r = 9.1$) but
not infinite.  The finite nature of this value is most likely due to
the limited overlapping spectral range ($\lambda$ =
2.2194--2.2290~$\mu$m) of the 2 observations of FU~Ori with different
instrumentation.  Imperfectly corrected instrumental differences may
have also contributed to reducing the significance of the
cross-correlation, and the intrinsic spectrum may have also changed
slightly between the two epochs.

\subsection{Near-IR Colors}

The near-IR $JHK$ colors of FUors and HHENs have not previously been
compared directly and these wavelengths encompass our spectra, so we
present a near-IR color-color diagram of 15 of these objects in
Figure~\ref{fig4} (including ones without spectra presented in this
paper).  The $H-K$ and $J-H$ colors in Figure~\ref{fig4} were all
computed from 2MASS catalog data, and the plot shows that most of the
HHENs have colors that are consistent with either reddened classical
T~Tauri stars (falling along the dashed T Tauri locus) or with
reddened normal stars with IR excesses; only HH354 IRS has colors that
are clearly inconsistent with IR excess.  FU~Ori, several FUors, and a
few FUor-like stars also have colors consistent with reddened
early-type stars (early-type stars are near the origin of the plot).
However, early-type stars are ruled out and the color degeneracy is
broken by the fact that the FUors and FUor-like objects generally show
strong near-IR CO absorptions, consistent with late spectral types
typical of T Tauri stars.  Therefore most of the FUors and FUor-like
HHENs have near-IR colors consistent with having circumstellar disks.
However, the large range of colors makes it difficult to differentiate
regular stars, FUors, and T~Tauri stars (or protostars) by their
colors alone.  This color range may be produced by some combination of
local scattering produced by different amounts and distributions of
circumstellar material as well as different object orientations (i.e.,
disk / envelope orientations).

\section{Discussion}

The similarity of the FUor and FUor-like spectra and their significant
cross-correlations indicate that the spectral lines of these objects
must form in physically similar environments.  \citet{RA97} had also
postulated this based on similarities of low resolution near-IR
spectra, and the high resolution data presented in this paper confirm
that the FUor-like HHENs are more similar to FUors than dwarfs,
giants, supergiants, or embedded protostars.  

\begin{figure}
\includegraphics[ scale=0.8,angle=0]{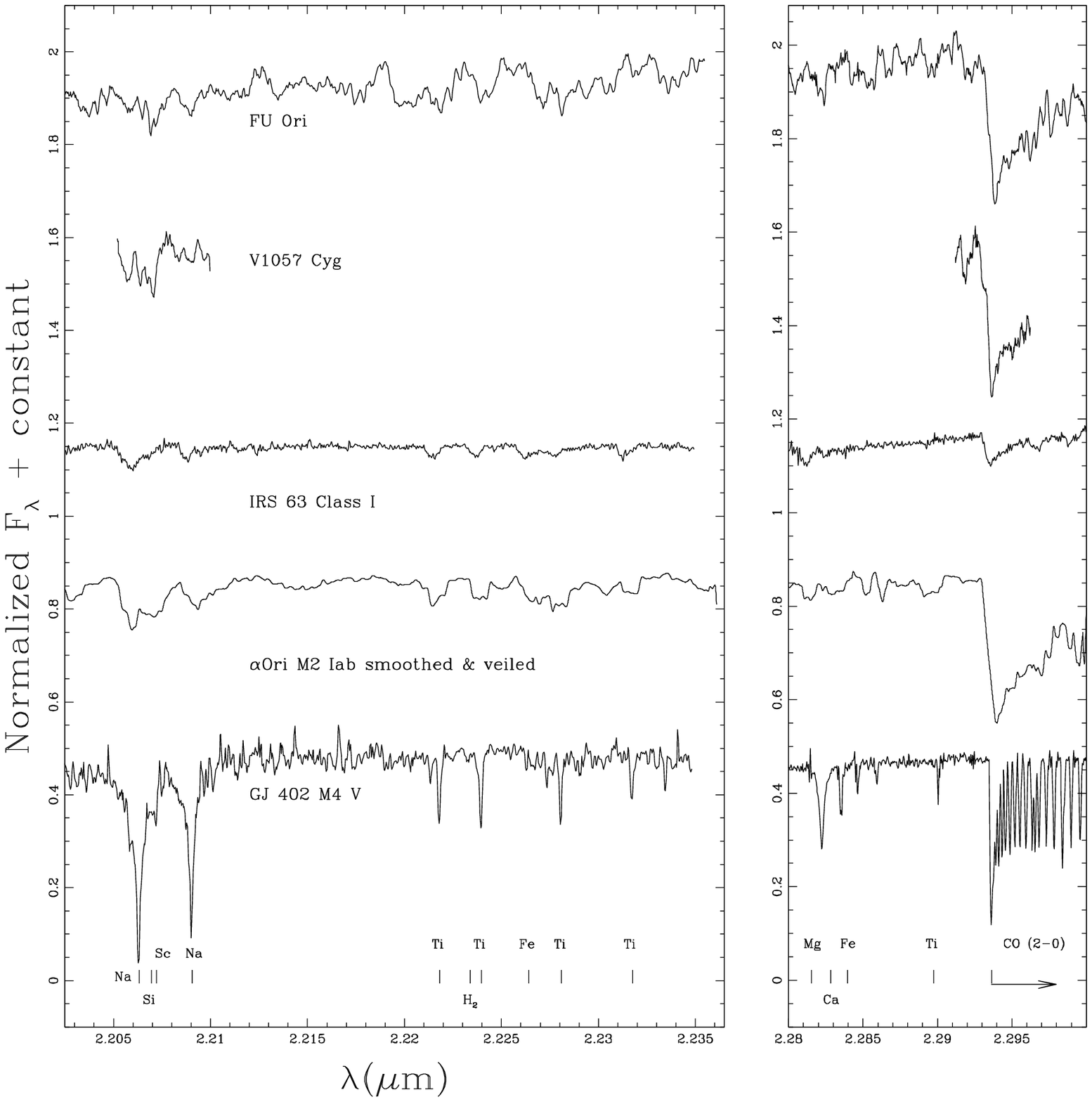}
\caption{Near-IR spectra of two FUors, a Class~I protostar, and
late-type stellar standards. The FUors do not strongly resemble the
other spectra except for having strong CO and some absorptions near
the Na lines.\label{fig1}}
\end{figure}

\begin{figure}
\includegraphics[ scale=0.8,angle=0]{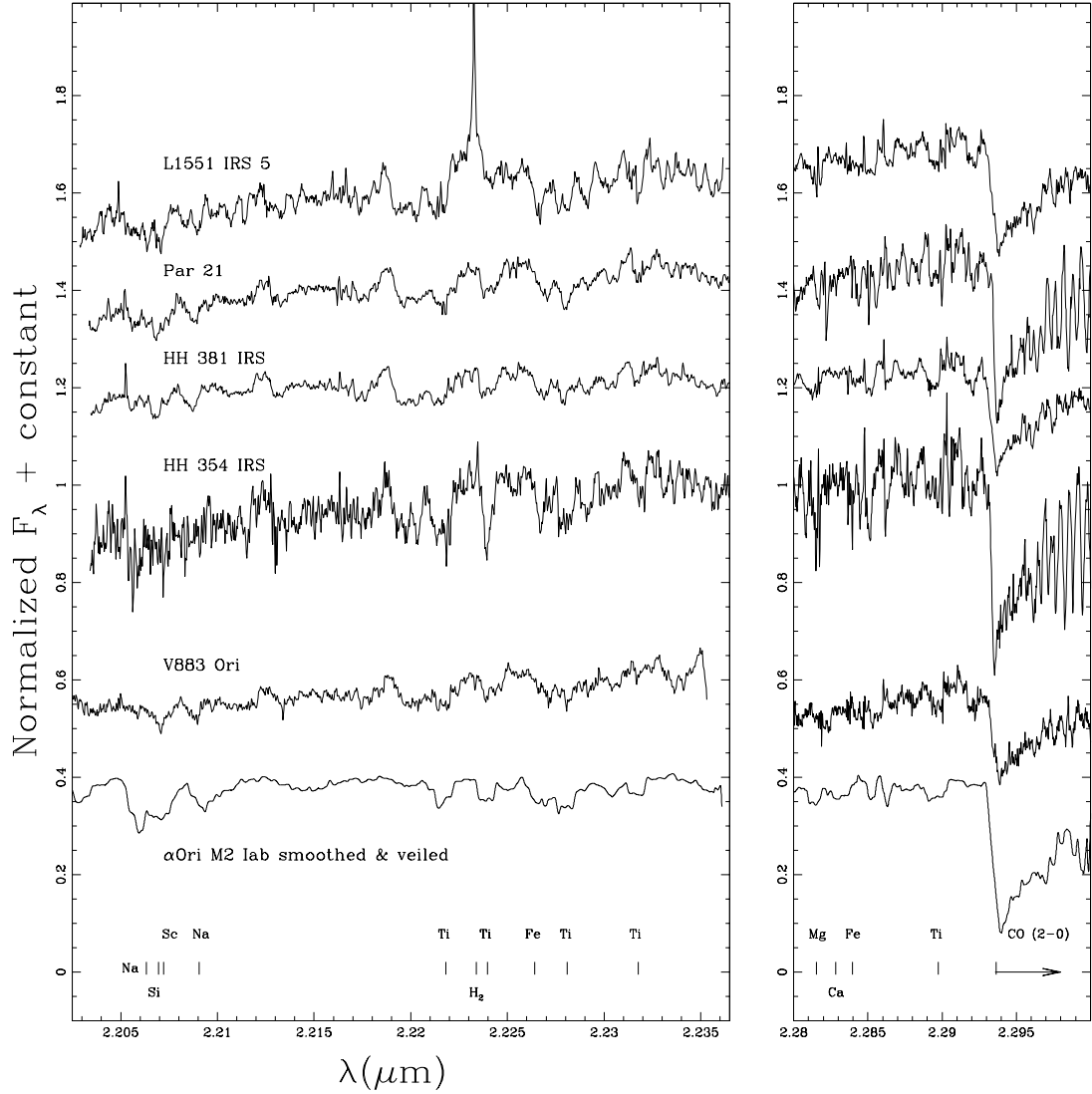}
\caption{Spectra of five FUor-like stars and the M2~Iab star
$\alpha$ Ori. The spectra of $\alpha$ Ori have been smoothed so that
the CO bandhead has approximately the same slope of the FUor-like
stars, $v$~sin~$i \simeq 100$ km~s$^{-1}$.\label{fig2}}
\end{figure}

\begin{figure}
\includegraphics[ scale=0.8,angle=0]{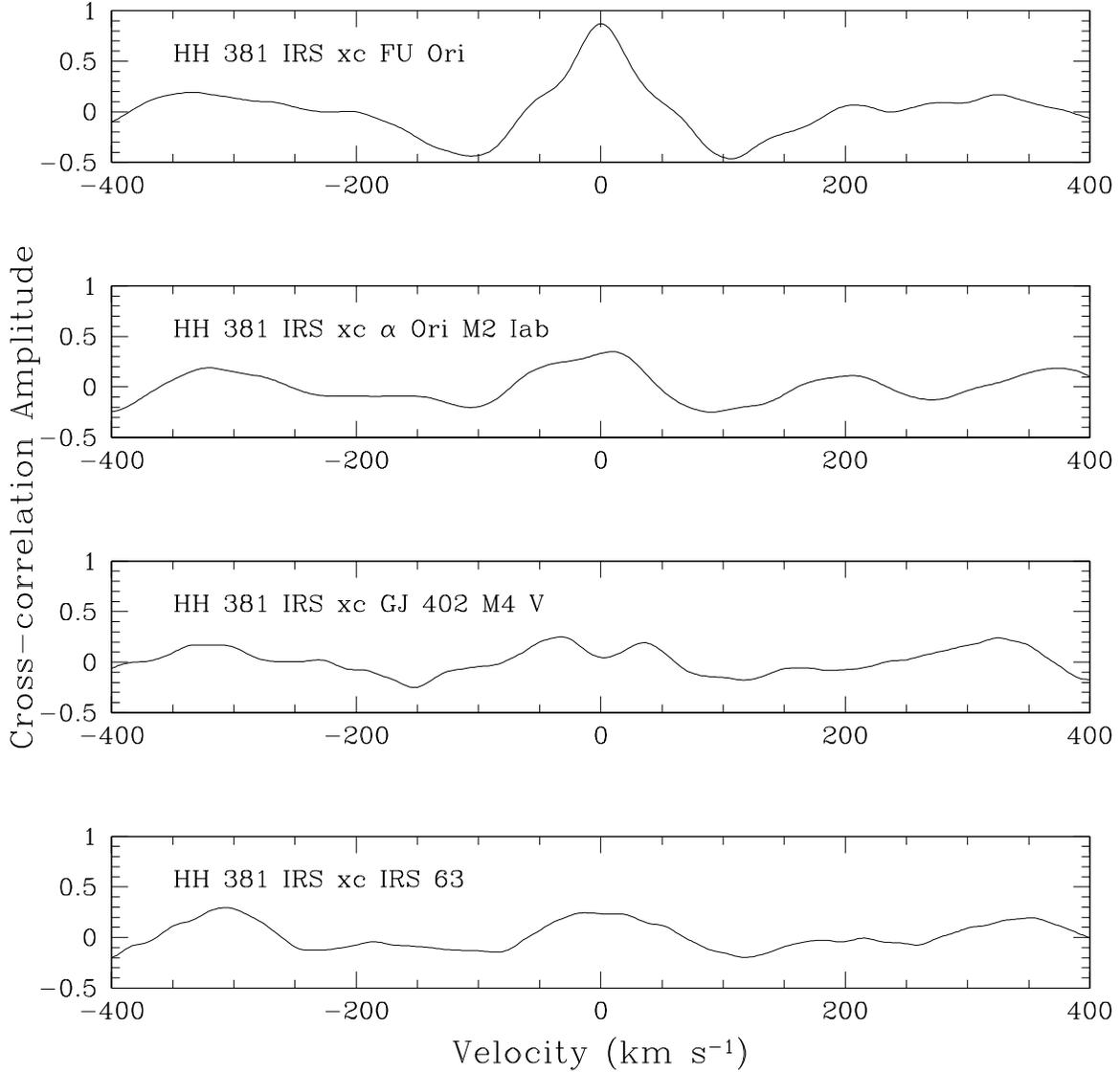}
\caption{Cross-correlation functions of the 2.203--2.236~$\mu$m
spectrum of the FUor-like star HH~381~IRS. The high central peak and
symmetrical nature of the cross-correlation in the top panel indicates
that the spectra of HH~381~IRS and FU~Ori have similar features, but
correlations with the giant, dwarf, and embedded protostar are poor.
Spectra have been shifted slightly in wavelength (and therefore
radial velocity) to produce symmetrical cross-correlation
peaks at 0 km~s$^{-1}$.\label{fig3}}
\end{figure}

Given the cross-correlation results, the near-IR spectra of FUor-like
stars do not appear to be produced in normal stellar photospheres.
The broad cross-correlation peaks indicate rotational velocities
$v$~sin~$i \sim 100$~km~s$^{-1}$, much larger than measured for
T~Tauri stars or even the most rapidly spinning embedded protostars
\citep{DGCL05, CGDL06}.  The cross-correlations with late-type dwarfs
and the embedded protostar IRS~63 are also generally poor (low
significance), but they are somewhat better with late-type giants and
supergiants (Table~\ref{tbl-2}).  This suggests that the spectral
features of FUor-like stars originate in cool regions with low surface
gravities that differ from normal stellar photospheres, and the
generally good cross-correlations with FUor spectra reinforce this.
The strong CO absorptions of the FUors and FUor-like stars
(Figures~\ref{fig1} and \ref{fig2}) indicate relatively low effective
temperatures, T$_{eff} \sim 3500$~K \citep[e.g.,][]{KH86}.  A giant or
supergiant star of that effective temperature would have a radius
greater than about 40 $R_{\odot}$ and a mass of approximately 6
$M_{\odot}$ or more.  Such stars have rotational breakup velocities on
the order of 100 km~s$^{-1}$.  Therefore, the CO strengths and
projected rotation velocities of FUor and FUor-like spectra (from
cross-correlation and CO bandhead widths; Fig.~\ref{fig2} and
\ref{fig3}) are consistent with giants rotating at breakup.  However,
the marginal cross-correlation significance with giants or supergiants
(Table~\ref{tbl-2}) indicates that the FUors and FUor-like stars are
not likely to be ordinary giants or supergiants.

\begin{figure}
\includegraphics[ scale=0.85,angle=0]{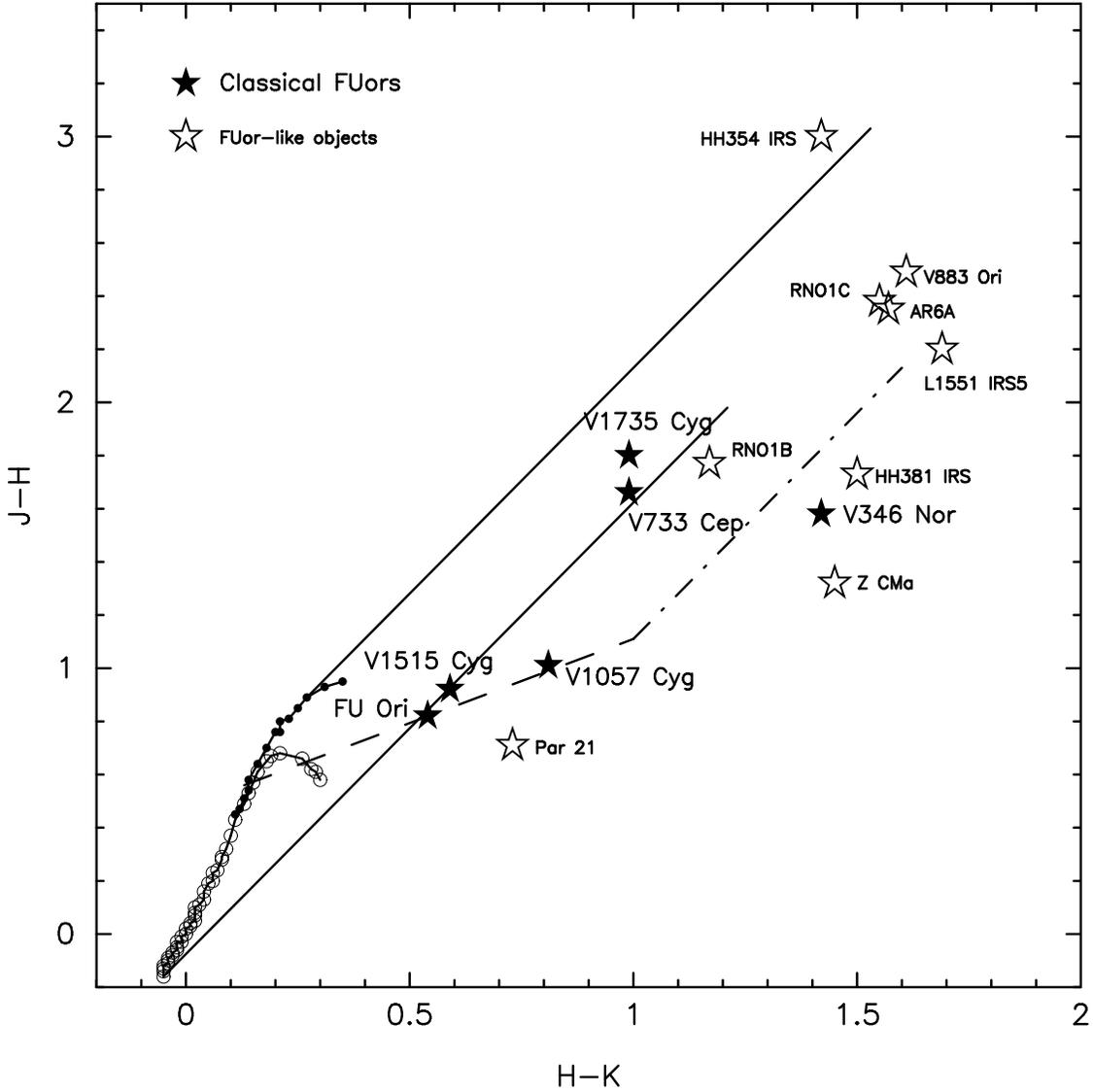}
\caption{Near-IR color-color plot of classical FUors (filled star
symbols) and FUor-like stars (open star symbols).  Several stars
posited to be FUor-like stars \citep[e.g., see][for
references]{AKCKMP04} but not otherwise analyzed in this work are
included.  $JHK$ magnitudes were obtained from the 2MASS database, and
uncertainties in colors are generally smaller than the symbol size.
Most FUors and FUor-like stars have near-IR colors that are
consistent with reddened T~Tauri stars with disks \citep[classical T
Tauri locus of][shown as the dashed line]{MCH97} but not reddened
late-type dwarfs (open circles and their locus) or giants (filled dots
and their locus).  The dot-dashed line indicates reddening equivalent
to $A_{v} = 10$~mag.\label{fig4}}
\end{figure}

The combined visible and IR spectra of FUors are reasonably well fit
by models of line formation in active circumstellar accretion disks
\citep[e.g.,][]{KHH88, HHC04}; the decrease in their line widths (and
implied rotational velocities) at longer wavelengths is recognized as
perhaps the best evidence of line formation in disks.  The
double-peaked nature of the cross-correlation of FUors and FUor-like
near-IR spectra with those of late type dwarfs (Figure~\ref{fig3}) is
also consistent with a disk rotational velocity profile, and the
significant near-IR excesses exhibited by most of these objects (e.g.,
$\S 4.3$ and Fig.~\ref{fig4}) are also consistent with hot
circumstellar disk material.

If the spectra of FUors and FUor-like HHENs are indeed produced in
accretion disks, then why are they not more similar to the spectra of
embedded protostars with active disks?  Even Class~I protostars with
large veiling and relatively high accretion rates $\dot{M} \simeq
10^{-6} M_{\odot}$~yr$^{-1}$ do not have FUor-like spectra \citep[see
also][]{GL02, DGCL05}.  In a circumstellar disk model atmosphere, high
accretion rates are needed to induce a temperature gradient with a hot
midplane and cooler exterior to produce a spectrum similar to a FUor
\citep[e.g.,][]{KHH88}.  \citet{CHS91} found that an irradiated
accretion disk model with mass accretion rate of $1.6 \times 10^{-4}
M_{\odot}$ yr$^{-1}$ reproduced the near-IR spectrum of FU~Ori well,
including its CO and H$_{2}$O absorptions.  This accretion rate is
approximately 2 orders of magnitude higher than that of embedded low
mass protostars such as IRS 63.  \citet{GL02} estimated that the mass
accretion rate of the embedded protostars YLW 15A was $2.3 \times
10^{-6} M_{\odot}$ yr$^{-1}$, and the same modeling would produce a
very similar (within $\sim50$\%) value for IRS 63 since the two
protostars have virtually identical effective temperatures, near-IR
veilings, rotational velocities \citep{DGCL05} and IRAS fluxes.  It
appears that higher accretion rates are needed to produce luminous
disks with absorption features similar to those found in FUors.
Future discoveries and observations of weaker FUors with lower
accretion rates and more active protostars with higher accretion rates
would be valuable in constraining further the accretion rate at which
a circumstellar disk develops a strong enough temperature gradient to
produce absorption lines that dominate those of the stellar
photosphere.


The very high degree of similarity of the FUor-like and FUor spectra
(and their good cross-correlations) suggest that FUor activity may be
more common than indicated from the small number of known classical
FUors alone; more young stars have spectra likely to originate in
luminous accretion disks.  Taking only the list of 20 classical and
FUor-like objects by \citet{AKCKMP04} and adding the two other HH
sources found to have FUor-like spectra by \citet{RA97} (HH~381~IRS
and HH~354~IRS; also confirmed in this work) yields a sample of 22
objects.  Of these, five are widely acknowledged to be bona fide
classical FUors due to their observed outbursts and spectral
properties (FU~Ori, V1057 Cyg, V1515 Cyg, V1735 Cyg, and V346 Nor).
Therefore FUor-like stars may outnumber FUors.

However, FUors are certainly not ubiquitous.  Very few FUors or stars
with FUor-like spectra are found in regions of clustered star
formation such as Tau-Aur, $\rho$~Oph, NGC~1333, IC~346, or the Orion
Nebula Cluster.  Instead, they are mostly found in regions of low star
formation activity.  This implies that either FUors are much
less frequent or have many fewer outburst cycles in high density
regions.  One possible explanation is that interactions between stars
in clusters could disrupt disks, perhaps eliminating much FUor
activity during the Class 0 or Class I protostellar evolutionary
phases.  Alternatively, FUor events may be triggered by the presence
of a companion star \citep{BB92}, in which case a dense cluster
environment might accelerate the orbital evolution, causing FUor
eruptions in clusters to occur mostly during the Class 0 or Class I
stages \citep{RA04}.  However, these are certainly not definitive
explanations, and the causes of FUor events must be understood better
before the inhomogeneity in the distribution of FUors and FUor-like
stars can be understood.

\acknowledgments

We thank G. Blake and C. Salyk for acquiring some of the Keck NIRSPEC
data and also thank K. Covey for reducing some NIRSPEC data.  We thank
G. Doppmann, P. Fukumura-Sawada, W. Golisch, D. Griep, K. Hinkle, G.
Puniwai, and C. Wilburn for assistance with the observations.  The
authors wish to recognize and acknowledge the very significant
cultural role and reverence that the summit of Mauna Kea has always
had within the indigenous Hawaiian community.  We are most fortunate
to have the opportunity to conduct observations from this mountain.
Observations of FU~Ori were obtained at the Gemini Observatory
(Program ID GS-2006A-C-12), which is operated by the Association of
Universities for Research in Astronomy, Inc., under a cooperative
agreement with the NSF on behalf of the Gemini partnership: the
National Science Foundation (United States), the Science and
Technology Facilities Council (United Kingdom), the National Research
Council (Canada), CONICYT (Chile), the Australian Research Council
(Australia), CNPq (Brazil) and CONICET (Argentina) This paper is based
on observations obtained with the Phoenix infrared spectrograph,
developed and operated by the National Optical Astronomy Observatory.
This publication makes use of data products from the Two Micron All
Sky Survey, which is a joint project of the University of
Massachusetts and the Infrared Processing and Analysis
Center/California Institute of Technology, funded by the National
Aeronautics and Space Administration (NASA) and the National Science
Foundation.  TPG acknowledges support from NASA's Origins of Solar
Systems program via WBS 811073.02.07.01.09.  This material is based
upon work supported by NASA through the NASA Astrobiology Institute
under cooperative agreement number NNA04CC08A, issued through the
Office of Space Science / Science Mission Directorate.  BR
acknowledges partial support for this study by National Science
Foundation grant AST-0407005.



\clearpage



\clearpage

\clearpage

\clearpage

\clearpage



\clearpage

\begin{deluxetable}{lrrlrrl}

\tablenum{1}
\tablecaption{Journal of Observations\label{tbl-1}}
\tablehead{
\colhead{Source}      & 
\colhead{$\alpha$(2000)} & \colhead{$\delta$(2000)} &
\colhead{Observed Date}  &
\colhead{Int. Time}   & \colhead{S/N} &
\colhead{Observatory} \\ [0.2 ex]
\colhead{}            & 
\colhead{\small (hh mm ss.s)} & \colhead{($\arcdeg$  $\arcmin$  
$\arcsec$)} &
\colhead{(UT)}            &
\colhead{\small(minutes)}   & \colhead{} &
\colhead{}
}
\tablecolumns{7}
\startdata
\sidehead{FUor-like HHENs:}
L1551 IRS5 & 04 31 34.1 & +18 08 05 & 2001 Nov 6 & 20.0 & 90 & Keck\\

V883~Ori & 05 38 18.1 & $-$07 02 27 & 2007 Mar 6 & 1.0 & 140 & Keck\\

Parsamian~21  & 19 29 00.7  & +09 38 39 & 2001 Jul 7 & 33.0 & 220 &
Keck\\

HH~381~IRS & 20 58 21.4 & +52 29 27 & 2001 Jul 7 & 21.0 & 190 & Keck\\

HH~354~IRS & 22 06 50.7 & +59 02 49 & 2001 Jul 7 & 21.0 & 60 & Keck \\

\sidehead{FUors:}
FU~Ori    & 05 45 22.4 & +09 04 12  & 2007 Mar 6 &  1.0 & 120 & Keck\\

          &            &            & 2006 Apr 3\tablenotemark{a}
	  &  4.0 & 160 & Gemini-S\\

V1057 Cyg & 20 58 53.7 & +44 15 29 & 1999 Aug 30\tablenotemark{b}  
& 13.0 & 300 & IRTF\\
          &            &           & 1996 Sep 2\tablenotemark{c} 
	  &  2.0 & 250 & IRTF\\

\enddata
\tablenotetext{a}{Single order spectrum of the 2.2194--2.2290~$\mu$m 
region with spectral resolution $R \sim 40,000$. This spectrum is not 
shown in the figures of this paper (as it is largely superseded by the
Keck spectra), but it is used in the cross-correlation analysis.}
\tablenotetext{b}{Spectrum of the 2.29353~$\mu$m 
$v$=0--2~CO bandhead region with spectral bandwidth $\sim 57$\AA.}
\tablenotetext{c}{ Spectrum of the 2.2075~$\mu$m Na line region with
spectral bandwidth is $\sim 55$\AA.  These spectra were originally 
published in Greene \& Lada (1997).}

\end{deluxetable}

\clearpage

\begin{deluxetable}{lrrrrr}

\tablenum{2}
\tablecaption{Cross-correlation $r$-values\label{tbl-2}}
\tablehead{
\colhead{Object}   & 
\colhead{FU~Ori}   & 
\colhead{$\alpha$ Ori} &
\colhead{HR 5150}  &
\colhead{GJ 402}   & 
\colhead{IRS 63} \\ [0.2 ex]
\colhead{} &
\colhead{} &
\colhead{M2 Iab} &
\colhead{M1.5 III} &
\colhead{M4 V} &
\colhead{protostar} 
}
\tablecolumns{6}
\startdata

L1551~IRS~5\tablenotemark{a} & 4.1 & 2.5 & 3.2 & 2.3 & 3.5 \\
V883~Ori & 28.7 & 4.3 & 4.0 & 3.8 & 3.8 \\
Parsamian~21 & 21.6 & 3.1 & 3.2 & 2.9 & 3.1\\
HH~381~IRS & 13.3 & 1.6 & 2.5 & 1.7 & 2.3 \\
HH~354~IRS & 7.5 & 3.0 & 3.0 & 1.9 & 3.3 \\
FU~Ori & $\infty$ & 3.4 & 2.9 & 3.3 & 3.1\\


\enddata
\tablenotetext{a}{The 2.223~$\mu$m H$_{2}$ emission line was
artificially removed from the spectrum of L1551 IRS~5 before computing
its cross-correlations.}

\end{deluxetable}

\begin{deluxetable}{lrrrrrr}
\tablenum{3}

\tablecaption{Cross-correlation Between Objects
($r$-values)\label{tbl-3}}
\tablewidth{0pt}
\tablehead{
\colhead{Object}   & 
\colhead{L1551~IRS~5\tablenotemark{a}}   & 
\colhead{V883~Ori}   &
\colhead{Parsamian~21}   &
\colhead{HH~381~IRS}   &
\colhead{HH~354~IRS}   &
\colhead{FU~Ori}   
}
\tablecolumns{7}
\startdata

L1551~IRS~5\tablenotemark{a} & $\infty$ & 5.0 & 6.0 & 6.6 &
8.1 & 5.1 \\
V883~Ori & 5.0 & $\infty$ & 25.2 & 18.8 & 8.7 & 28.7\\
Parsamian~21 & 6.0 & 25.2 & $\infty$ & 21.5 & 9.6 & 21.6\\
HH~381~IRS & 6.6 & 18.8 & 21.5 & $\infty$ & 6.7 & 13.3\\
HH~354~IRS & 8.1 & 8.7 & 9.6 & 6.7 & $\infty$ & 7.5 \\
FU~Ori & 5.1 & 28.7 & 21.6 & 13.3 & 7.5 & 9.1\tablenotemark{b}\\

\enddata 
\tablenotetext{a}{The 2.223~$\mu$m H$_{2}$ emission line was
artificially removed from the spectrum of L1551 IRS~5 before computing
its cross-correlations.} 
\tablenotetext{b}{The 2007 Mar 6 Keck NIRSPEC
spectrum of FU~Ori, shown in Figure 1 and used for all FU~Ori
cross-correlations, was cross-correlated with a spectrum of FU~Ori
acquired on 2006 April 03 with the GEMINI-S telescope (See Table 1).}
\end{deluxetable}




\end{document}